\documentclass{article}
\usepackage{spconf,amsmath,epsfig}

\let\OLDthebibliography\thebibliography
\renewcommand\thebibliography[1]{
  \OLDthebibliography{#1}
  \setlength{\parskip}{0pt}
  \setlength{\itemsep}{0pt plus 0.3ex}
}

\usepackage{url}
\usepackage{graphicx}
\usepackage{amsmath}
\usepackage{amsthm}
\usepackage{subcaption}
\usepackage{enumitem}
\usepackage{booktabs}
\usepackage{multirow}
\usepackage{amsfonts,amssymb}
\usepackage{hyperref}
\hypersetup{hidelinks}

\def\equationautorefname#1#2\null{
	Eq.(#2\null)
}

\begin{document}\sloppy

\def\x{{\mathbf x}}
\def\L{{\cal L}}

\title{Jointly Recognizing Speech and Singing Voices Based on Multi-Task Audio Source Separation}
%

\name{Ye Bai$^{1*}$, Chenxing Li$^{1,2*}$, Hao Li$^1$, Yuanyuan Zhao$^1$, Xiaorui Wang$^1$\thanks{* denotes equal contribution.}}
\address{
  $^1$Institute of Automation, Chinese Academy of Sciences, Beijing, China\\
  $^2$Tencent AI Lab, Beijing, China\\
}

\maketitle

\begin{abstract}
In short video and live broadcasts, speech, singing voice, and background music often overlap and obscure each other. This complexity creates difficulties in structuring and recognizing the audio content, which may impair subsequent ASR and music understanding applications. This paper proposes a multi-task audio source separation (MTASS) based ASR model called JRSV, which \textbf{J}ointly \textbf{R}ecognizes \textbf{S}peech and singing \textbf{V}oices. Specifically, the MTASS module separates the mixed audio into distinct speech and singing voice tracks while removing background music. The CTC/attention hybrid recognition module recognizes both tracks. Online distillation is proposed to improve the robustness of recognition further. To evaluate the proposed methods, a benchmark dataset is constructed and released. Experimental results demonstrate that JRSV can significantly improve recognition accuracy on each track of the mixed audio.
\end{abstract}
\section{Introduction}
\label{sec:intro}

The audio signals recorded in live broadcasts and short videos usually contain various types of sources, such as speech, singing voices, background music, and sound effects. These signals often overlap and obscure each other, which increases the difficulty of speech and lyrics recognition, and the accuracy drops significantly. In such scenarios, speech and singing voices are equally important. The accurate recognition both contribute to subsequent recommendation systems and search engines. It is necessary to improve the accuracy of speech and lyrics recognition in this overlapping scenarios. 

Conventionally, recognizing multi-track of speech from monaural audio uses cascade systems \cite{chen2020continuous}, i.e., a speech separation system is first used to separate the speech of multiple talkers from the mixed audio, and then an ASR system recognizes the content of each track. However, the mismatch between the separated audio and the natural audio hurts the recognition performance of the system. Moreover, previous separation models, such as deep clustering \cite{hershey2016deep}, permutation invariant training (PIT) \cite{kolbaek2017multitalker}, TasNet \cite{luo2019conv}, do not distinguish the type of the separated tracks. Thus the user cannot distinguish the content of the speech and the singing voices. Another type of method to recognize the mixed audio is the end-to-end methods, such as extended PIT \cite{yu17b_interspeech,settle2018end,chang2019end,lu2021streaming,raj2022continuous} and serialized output training (SOT) \cite{kanda20b_interspeech}. These methods optimize the model in an end-to-end way and show good performance. However, these methods also cannot distinguish the type of the recognized tracks. Besides, PIT-based methods also meet permutation problem, which may make ASR models confuse.

To achieve a detailed structured result of the speech and singing voices that mixed in audio, in this paper, we propose a unified model to Jointly Recognize Speech and singing Voices (JRSV). JRSV provides the types of audio tracks and recognizes the content of the speech and singing voices. This is the first time to investigate how to jointly separate and recognize speech and singing voices in overlapping scenes. The contributions are listed: (1) We divide JRSV into two modules: a multi-task audio source separation (MTASS) module and an ASR module. The MTASS module separates the mixed audio into a speech track and a singing voice track. It also removes the background music at the same time. Due to PIT-free in MTASS, JRSV avoids the permutation and selection problems. Then the ASR module recognizes the content of the two tracks. (2) We adopt two-stage training and employ online distillation to make the encoded representations of separated tracks to approximate the representations of the clean audio track to improve the robustness of the model. (3) To evaluate the proposed methods, we build and release a benchmark dataset called Dual-Track Speech and singing Voice Dataset (DTSVD). The experimental results demonstrate that JRSV outperforms the cascade system by achieving a relative reduction of 41\% in character error rates (CERs) for speech and 57\% in CER for singing voices.

\section{Related work}

\textbf{Multi-speaker speech separation.} Speaker-independent speech separation aims to separate the speech of different speakers. Frequency domain-based DPCL \cite{hershey2016deep}, PIT \cite{kolbaek2017multitalker}, CBLDNN-GAT \cite{li2018cbldnn}, and time domain-based TasNet \cite{luo2019conv}, DPRNN \cite{luo2020dual} have achieved state-of-the-art (SOTA) performance. However, PIT-based methods meet several challenges: an unknown number of sources in the mixture, permutation problem, and selection from multiple outputs.

\textbf{MTASS.} MTASS aims to separates speech, music, and background noise simultaneously \cite{zhang2021multi,petermann2022cocktail,li2022ead, wang2022wa, petermann2023tackling}. Without PIT, MTASS avoids permutation problem. This work extends the MTASS to separate speech and singing voices and unify the model to the end-to-end ASR system. The model can separate and recognize the speech and the singing voice track while removing the mixture's music. 

\textbf{Speech recognition and lyrics recognition.} Recently, deep learning has brought much progress in speech recognition \cite{li2022recent}. As a special type of speech, lyrics recognition \cite{gao2022automatic,dabike2019automatic} also draws the attention of speech researchers. 
\cite{petermann2023tackling} first adopts MTASS to extract speech and music. But it after only focuses on speech recognition.
Conventionally, these two topics are discussed independently, and a real-world situation, i.e., the speech and the singing voices are mixed in one channel, is not considered. Different from previous work, this paper provides a model to jointly recognize speech and singing voices.

\textbf{Multi-talker ASR.} Previously, PIT \cite{yu17b_interspeech,lu2021streaming,raj2022continuous,zhang2020improving} and SOT \cite{kanda20b_interspeech}-based end-to-end methods are proposed for multi-talker ASR. However, these methods cannot provide the track label and the detailed structure of the recognition results. Different from these works, this work employs a direct way to separate the speech track and singing voice track and then recognize each track, which is suitable for analyzing the structure of the recognized tracks.

\begin{figure}[t]
  \centering
  \includegraphics[width=0.8\linewidth]{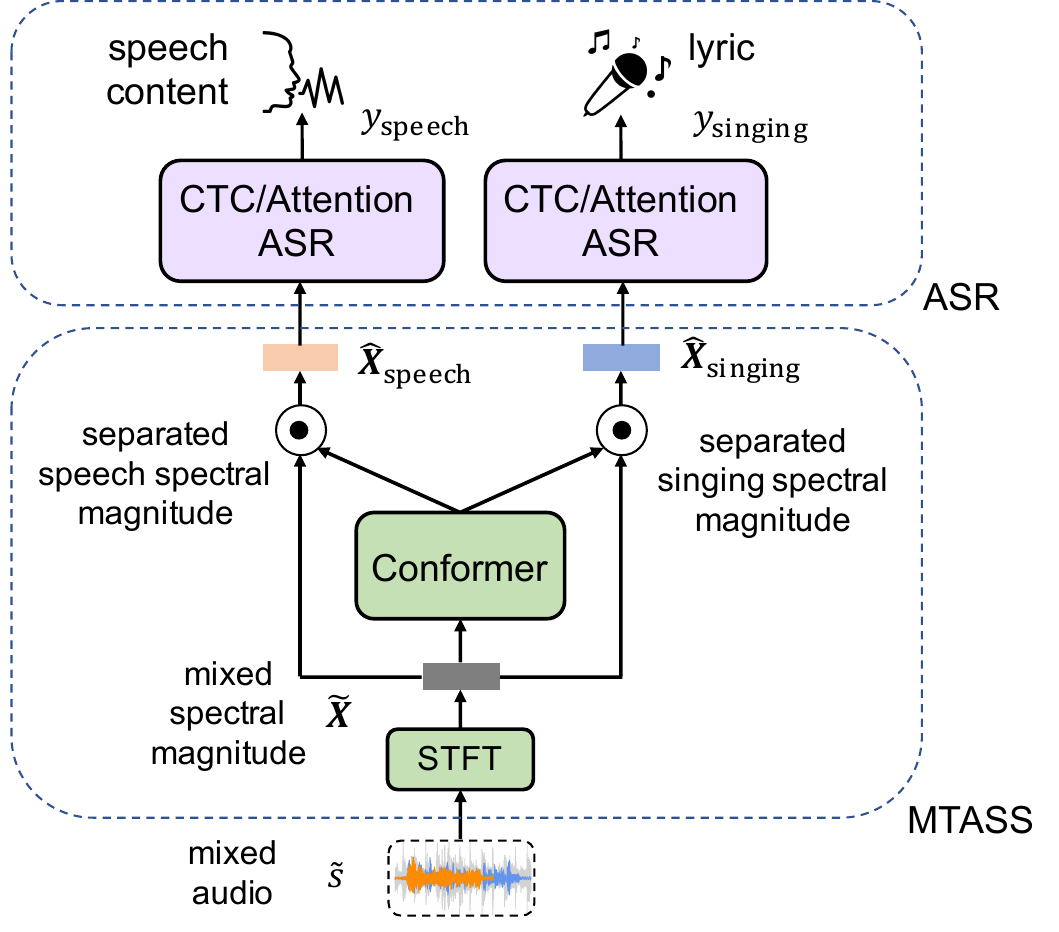}
  \caption{An overview of JRSV system. $\widetilde{\mathbf{s}}$ denotes the mixed audio wave. $\widetilde{\mathbf{X}}$ denotes the mixed spectral magnitude. $\hat{\mathbf{X}}_{\text{speech}}$ and $\hat{\mathbf{X}}_{\text{singing}}$ denote the separated spectral magnitudes of speech and the singing voice respectively. $y_{\text{speech}}$ and $y_{\text{singing}}$ denote the text sequence of speech and singing voice.}
  \label{fig:sys}
  \vspace{-10pt}
\end{figure}

\section{Proposed methods}

This section introduces the proposed JRSV and then describes each module. \autoref{fig:sys} shows the overview structure. The system consists of two parts. The MTASS module separates the mixed audio into a speech track and a singing voice track. The ASR module recognizes the content of the speech track and singing voice track.

\subsection{The MTASS module}

We adopt the Conformer-based \cite{li2022ead, gulati20_interspeech} MTASS network to separate speech and singing voices. In detail, after a short-time Fourier transform (STFT), the spectral magnitude is fed into the Conformer-based separation network. In the last layer, the separated spectral magnitude of speech and singing voices are mapped out using two output layers. 

\subsection{The ASR module} 
We employ the U2 (CTC/attention rescoring) structure \cite{yao21_interspeech,tian2021one} as the ASR model. The model consists of three parts. Given the separated spectral magnitude, the Conformer-based encoder encodes the acoustic representations. Then, the CTC decoder uses prefix beam search to generate candidates. And then, the attention decoder rescores the candidates to find the best hypothesis. As transformer-based attention rescorer can be executed in parallel, the rescoring process runs fast.


\subsection{Training}
\subsubsection{MTASS losses}
Three kinds of losses are applied to MTASS: magnitude-based separation loss, discriminate separation loss, and consistency loss. Magnitude-based separation loss aims to minimize the energy between the clean source and the estimated source. Discriminate separation loss measures the difference between the estimated source and other sources, which seeks to enhance the discriminative power of the model. The consistency loss is computed between the mixture and the sum of the estimated source. These losses are based on utterance-level $L_1$ norm. The loss function is formulated as follows.
\begin{equation}
\begin{aligned}
&L_{\text{MTASS}} = L_{\text{mag}} - \lambda L_{\text{dis}} + \gamma L_{\text{cst}}, \\
&L_{\text{mag}} = || \hat{\mathbf{X}}_\text{speech} - {\mathbf{X}}_\text{speech} || + || \hat{\mathbf{X}}_\text{sing} - {\mathbf{X}}_\text{sing} ||, \\
&L_{\text{dis}} = || \hat{\mathbf{X}}_\text{speech} - {\mathbf{X}}_\text{sing} || + || \hat{\mathbf{X}}_\text{sing} - {\mathbf{X}}_\text{speech} ||, \\
&L_{\text{cst}} = || (\hat{\mathbf{X}}_\text{speech} + \hat{\mathbf{X}}_\text{sing})- ( {\mathbf{X}}_\text{speech} + {\mathbf{X}}_\text{sing}) || ,
\end{aligned}
\end{equation}
where $\lambda$ and $\gamma$ are scale parameters to balance each loss. $\hat{\mathbf{X}}_\text{speech}$, $\hat{\mathbf{X}}_\text{sing}$, $\mathbf{X}_\text{speech}$ and $\mathbf{X}_\text{sing}$ denote the spectral magnitude of the separated speech, separated singing voices, clean speech, and clean singing voices, respectively. In the experiments, we set $\lambda$ to 0.1 and $\gamma$ to 0.3.

\begin{table*}[t]
\caption{Preliminary experiment: ASR models directly recognize the test sets of DTSSVD without separation. \texttt{speech} means the model trained with only speech data. \texttt{singing} means the model trained with only singing data. \texttt{multi} means the modeled trained with both speech and singing data.}
\label{tab:pri_exp}
\centering
\vspace{-10pt}
\begin{tabular}{|l|rrrrrr|}
\hline
\multirow{2}{*}{} & \multicolumn{6}{c|}{CER\% (Speech / Singing Voices)} \\ \cline{2-7} 
 &
  \multicolumn{1}{c|}{Unmixed} &
  \multicolumn{1}{c|}{Overlap 0.0} &
  \multicolumn{1}{c|}{Overlap 0.1} &
  \multicolumn{1}{c|}{Overlap 0.3} &
  \multicolumn{1}{c|}{Overlap 0.5} &
  \multicolumn{1}{c|}{Overlap 1.0} \\ \hline\hline
\texttt{ASR-speech} &
  \multicolumn{1}{r|}{6.3 / 88.2} &
  \multicolumn{1}{r|}{48.1 / 245.7} &
  \multicolumn{1}{r|}{49.4 / 233.5} &
  \multicolumn{1}{r|}{58.5 /186.2} &
  \multicolumn{1}{r|}{69.1 / 169.1} &
  75.3 / 159.7 \\ \hline
\texttt{ASR-singing} &
  \multicolumn{1}{r|}{86.6 / 9.7} &
  \multicolumn{1}{r|}{185.5 / 95.6} &
  \multicolumn{1}{r|}{206.3 / 94.3} &
  \multicolumn{1}{r|}{167.6 / 89.5} &
  \multicolumn{1}{r|}{150.8 / 90.6} &
  145.1 / 91.6 \\ \hline
\texttt{ASR-multi} &
  \multicolumn{1}{r|}{6.3 / 9.4} &
  \multicolumn{1}{r|}{57.9 / 222.4} &
  \multicolumn{1}{r|}{60.3 / 236.9} &
  \multicolumn{1}{r|}{67.9 / 190.9} &
  \multicolumn{1}{r|}{76.2 / 171.8} &
  80.6 / 163.8 \\ \hline
\end{tabular}
\vspace{-10pt}
\end{table*}


\subsubsection{ASR losses}
The ASR module is trained with CTC/attention joint loss:
\begin{equation}\label{eq:loss_asr}
    L_{\text{ASR}} = \alpha L_{\text{CTC}} + (1-\alpha) L_{\text{att}},
\end{equation}
where $L_{\text{CTC}}$ is computed with a forward-backward algorithm \cite{graves2006connectionist} for the CTC decoder, and $L_{\text{att}}$ is computed with token-wise cross-entropy for the attention rescorer.

Inspired by previous robust ASR work \cite{li2017large,yi2018distilling,narayanan2022mask}, we propose online distillation to further improve the robustness of the representation ability of the Conformer encoder. Specifically, we attempt to make the encoded acoustic representations of the separated spectral magnitude approximate the encoded representations of the original clean spectral magnitude. 

Let $\hat{\mathbf{E}}$ denotes the encoded acoustic representations by the Conformer, and $\mathbf{E}$ is the encoded acoustic representations of the corresponding original clean audio. The distillation is as follows:
\begin{equation}\label{eq:loss_dist}
    L_{\text{distil}} = || \hat{\mathbf{E}}_\text{speech} - sg(\mathbf{E}_\text{speech}) || + || \hat{\mathbf{E}}_\text{sing} - sg(\mathbf{E}_\text{sing}) ||,
\end{equation}
where $sg(\cdot)$ is the stop gradient operation.

The final loss is as follows:
\begin{equation}
    L_{\text{ASR}} = \alpha L_{\text{CTC}} + (1-\alpha) L_{\text{att}} + \beta L_{\text{distil}},
\end{equation}
where the $\alpha$ and $\beta$ are the hyperparameters to balance the values during training. We set $\alpha$ to 0.3 and $\beta$ to 0.001.

The training procedure is as follows:
\begin{enumerate}
    \item Forward the original spectral magnitude of speech $\mathbf{X}_\text{speech}$ to the ASR module, and obtain acoustic representation $\mathbf{E}_\text{speech}$. Also, forward the original spectral magnitude of singing voices $\mathbf{X}_\text{sing}$ and obtain $\mathbf{E}_\text{sing}$.
    \item Forward the separated spectral magnitude of speech $\hat{\mathbf{X}}_\text{speech}$ to obtain the corresponding $\hat{\mathbf{E}}_\text{speech}$. Also, forward $\hat{\mathbf{X}}_\text{sing}$ and obtain $\hat{\mathbf{E}}_\text{sing}$.
    \item Compute \autoref{eq:loss_asr} for the original and separated spectral magnitude. (multi-condition training)
    \item Compute \autoref{eq:loss_dist} and optimize the parameters by back-propagation. (online distillation)
\end{enumerate}
Note that the original clean signal is only used to provide teacher supervision, and the gradient graph is truncated while computing the distillation losses.

\begin{table}[t]
  \caption{The statistics of the datasets.}
  \label{tab:dataset_info}
  \centering
\vspace{-10pt}
\begin{tabular}{|l|l|lc|}
\hline
\multicolumn{1}{|c|}{Dataset} & \multicolumn{1}{c|}{Type} & \multicolumn{2}{c|}{\#Dur.}         \\ \hline\hline
\multirow{3}{*}{AISHELL-1}  & \multirow{3}{*}{Speech}         & \multicolumn{1}{l|}{Training} & 150h \\ \cline{3-4} 
                              &                           & \multicolumn{1}{l|}{Dev.}     & 18h \\ \cline{3-4} 
                              &                           & \multicolumn{1}{l|}{Test}     & 10h \\ \hline
\multirow{3}{*}{OpenSinger} & \multirow{3}{*}{Singing Voices} & \multicolumn{1}{l|}{Training} & 34h  \\ \cline{3-4} 
                              &                           & \multicolumn{1}{l|}{Dev.}     & 10h \\ \cline{3-4} 
                              &                           & \multicolumn{1}{l|}{Test}     & 6h  \\ \hline
\multirow{3}{*}{MusDB18}      & \multirow{3}{*}{Music}    & \multicolumn{1}{l|}{Training} & 28h \\ \cline{3-4} 
                              &                           & \multicolumn{1}{l|}{Dev.}     & 7h  \\ \cline{3-4} 
                              &                           & \multicolumn{1}{l|}{Test}     & 19h \\ \hline
\multirow{3}{*}{DTSSVD}      & \multirow{3}{*}{Mixed}    & \multicolumn{1}{l|}{Training} & on-the-fly \\ \cline{3-4} 
                              &                           & \multicolumn{1}{l|}{Dev.}     & 18h  \\ \cline{3-4} 
                              &                           & \multicolumn{1}{l|}{Test}     & 10h \\ \hline

\end{tabular}
\vspace{-10pt}
\end{table}

\subsubsection{Two-stage training}
Counter-intuitively, we find that optimizing the $L_{\text{MTASS}}$ and $L_{\text{ASR}}$ at the same time does not improve the performance of the final ASR. Even the whole system cannot converge when the model is trained from scratch. We analyze these two reasons: 1) The spectral magnitude of the mixed audio contains ambiguous information, which makes the optimization procedure hard. 2) the high-level $L_{\text{ASR}}$ is not compatible with the low-level $L_{\text{MTASS}}$. To address this, we propose a curriculum learning \cite{bengio2009curriculum}-based two-stage training procedure. First, we train the MTASS module. Then, the ASR model is trained, and the MTASS module is fixed during training. We will discuss the reason why joint training does not improve performance in \autoref{subsec:exp_jrsv}.

\subsection{Inference}
As shown in \autoref{fig:sys}, during inference, the input audio is first separated into two tracks (the speech track and the singing voice track) by the MTASS module. Then each track is recognized by the ASR module. The two tracks can be recognized in parallel.

\section{DTSSV dataset} 
\label{sec:dataset}
DTSSV dataset is built based on three publicly available datasets: AISHELL-1\footnote{https://www.openslr.org/33/} \cite{bu2017aishell}, OpenSinger\footnote{https://github.com/Multi-Singer/Multi-Singer.github.io} \cite{huang2021multi}, and MusDB18\footnote{https://sigsep.github.io/datasets/musdb.html}\cite{musdb18}. AISHELL-1 is a Chinese ASR dataset that is recorded in reading form. OpenSinger is a Chinese singing voice dataset. MusDB18 is a multi-genre music dataset. In MusDB18, only background music is adopted. By combining and mixing these data, we built the DTSSV dataset. The basic information is listed in \autoref{tab:dataset_info}.

We randomly sample speech audio, a singing voice, and a background music segment and mix them with different overlaps. Specifically, we use the following mixing strategy.
\begin{enumerate}
    \item Normalize the amplitude of the speech $\mathbf{s}_{\text{speech}}$, singing voice $\mathbf{s}_{\text{sing}}$, and background music $\mathbf{s}_{\text{music}}$.
    \item Draw signal-to-noise ratios (SNRs) in decibels from uniform distributions for the three signals. We choose $\mathcal{U}(-10, 2)$ for $\mathbf{s}_{\text{speech}}$ and $\mathbf{s}_{\text{sing}}$, and $\mathcal{U}(-15, 2)$ for $\mathbf{s}_{\text{music}}$. Then the normalized signals are scaled with respect to the sampled SNRs.
    \item Mix the singing voice $\mathbf{s}_{\text{sing}}$ and the background music $\mathbf{s}_{\text{music}}$. Then randomly sample an overlap ratio from \{1.0, 0.5, 0.3, 0.1, 0.0\} and mix $\mathbf{s}_{\text{speech}}$ with the mixed singing voice.
\end{enumerate}

For the development set and the test set, we randomly select a singing voice and a music segment for each speech audio. For the training set, the three audio sources are mixed on-the-fly during training.

\section{Evaluation}
We evaluate the system in two aspects: separation performance and recognition accuracy. Signal-to-distortion ratio (SDR) \cite{sdr} improvements are used to evaluate the separation performance.

For recognition accuracy, we compute the CERs for the speech and singing voice tracks. Note that because the number of utterances of the singing voice is less than the speech, two mixed audio will contain the same singing voice track. To make the CERs of the singing voice track of mixed audio comparable with the CER of the original OpenSinger dataset, we fix the utterance list of the mixed audio and ensure that each singing voice exhibits only once.

\section{Experiments and analysis}
\subsection{Setup}

For MTASS, we use the Conformer-based model \cite{gulati20_interspeech}. The model consists of 16 Conformer blocks. The dimension of each block ($d_\text{model}$) is 256. The number of heads of the multi-head attention is 8. Relative positional encodings are used. The inner dimension of MLP ($d_\text{ffn}$) is 1024. The kernel size of the convolution is 33. The window length of the input wave is 1024, and the window stride is 256. The window size of the FFT is also 256. Thus, the dimension of the output spectral magnitude of the MTASS model is 513. We name the model \texttt{MTASS} in this paper. We use Adam optimizer \cite{kingma2014adam}, and the learning rate is set to 1e-3. All the MTASS models are trained for 200 epochs. The training samples are mixed on-the-fly with the same strategy in \autoref{sec:dataset}.

For CTC/attention hybrid ASR, we follow the configuration of wenet\footnote{https://github.com/wenet-e2e/\\wenet/blob/main/examples/aishell/s0/conf/train\_conformer.yaml}. The model first uses a 2 dimensional CNN for subsampling. 12 Conformer blocks are stacked as the encoder, and 6 Transformer decoder blocks are stacked as the rescorer. The dimension of each block is 256. The number of heads of the multi-head attention is 4. The inner dimension of MLP ($d_\text{ffn}$) is 2048. And the kernel size of the convolution is 15. For the cascade system, we use the 80-dim FBANK features. And for JRSV, the input dimension is 513. The optimizer is Adam. And we use the Noam learning rate schedule \cite{vaswani2017attention} with 10000 warm-up steps. All the ASR models are trained for 100 epochs. And for the two-stage training for JRSV, we first load the pre-trained MTASS and then train the ASR module.

\begin{table}[t]
  \caption{The SDR improvments of MTASS models on the test sets with different overlap ratios.}
  \vspace{-10pt}
  \label{tab:sep_exp}
  \centering
   \begin{tabular}{|ll|rrrrr|}
    \hline
    \multicolumn{2}{|l|}{} &
      \multicolumn{5}{c|}{SDRi} \\ \hline
    \multicolumn{2}{|l|}{Overlap Ratios} &
      \multicolumn{1}{c|}{0.0} &
      \multicolumn{1}{c|}{0.1} &
      \multicolumn{1}{c|}{0.3} &
      \multicolumn{1}{c|}{0.5} &
      \multicolumn{1}{c|}{1.0} \\ \hline\hline
    \multicolumn{1}{|l|}{} &
      speech &
      \multicolumn{1}{r|}{{ 46.0}} &
      \multicolumn{1}{r|}{{ 26.7}} &
      \multicolumn{1}{r|}{{ 18.0}} &
      \multicolumn{1}{r|}{{ 15.9}} &
      { 14.6} \\ \cline{2-7} 
    \multicolumn{1}{|l|}{\multirow{-2}{*}{\texttt{MTASS}}} &
      singing &
      \multicolumn{1}{r|}{{ 19.4}} &
      \multicolumn{1}{r|}{{ 18.0}} &
      \multicolumn{1}{r|}{{ 15.4}} &
      \multicolumn{1}{r|}{{ 14.4}} &
      { 13.7} \\ \hline
    \multicolumn{1}{|l|}{} &
      speech &
      \multicolumn{1}{r|}{{ 44.8}} &
      \multicolumn{1}{r|}{{ 27.0}} &
      \multicolumn{1}{r|}{{ 18.5}} &
      \multicolumn{1}{r|}{{ 16.5}} &
      { 15.1} \\ \cline{2-7} 
    \multicolumn{1}{|l|}{\multirow{-2}{*}{+ ${L}_{dis}$}} &
      singing &
      \multicolumn{1}{r|}{{ 19.1}} &
      \multicolumn{1}{r|}{{ 17.6}} &
      \multicolumn{1}{r|}{{ 15.0}} &
      \multicolumn{1}{r|}{{ 14.1}} &
      { 13.3} \\ \hline
    \multicolumn{1}{|l|}{} &
      speech &
      \multicolumn{1}{r|}{{ \textbf{49.0}}} &
      \multicolumn{1}{r|}{{ \textbf{27.3}}} &
      \multicolumn{1}{r|}{{ \textbf{18.7}}} &
      \multicolumn{1}{r|}{{ \textbf{16.6}}} &
      { \textbf{15.4}} \\ \cline{2-7} 
    \multicolumn{1}{|l|}{\multirow{-2}{*}{+ ${L}_{cst}$}} &
      singing &
      \multicolumn{1}{r|}{{ \textbf{19.7}}} &
      \multicolumn{1}{r|}{{ \textbf{18.2}}} &
      \multicolumn{1}{r|}{{ \textbf{15.6}}} &
      \multicolumn{1}{r|}{{ \textbf{14.6}}} &
      { \textbf{14.0}} \\ \hline
    \end{tabular}
    \vspace{-10pt}
\end{table}

\begin{table*}[t]
\caption{The CERs of the cascade system on the test sets with different overlap ratios.}
\label{tab:cascade_exp}
\centering
\vspace{-10pt}
\begin{tabular}{|l|rrrrrr|}
\hline
 & \multicolumn{6}{c|}{CER \% (Speech / Singing Voices)} \\ \hline
\multicolumn{1}{|c|}{Overlap Ratios} &
  \multicolumn{1}{c|}{0.0} &
  \multicolumn{1}{c|}{0.1} &
  \multicolumn{1}{c|}{0.3} &
  \multicolumn{1}{c|}{0.5} &
  \multicolumn{1}{c|}{1.0} &
  \multicolumn{1}{c|}{Avg.} \\ \hline\hline
\texttt{ASR-multi} / \texttt{MTASS} &
  \multicolumn{1}{r|}{9.2 / 28.4} &
  \multicolumn{1}{r|}{12.8 / 29.6} &
  \multicolumn{1}{r|}{24.3 / 32.5} &
  \multicolumn{1}{r|}{29.3 / 35.9} &
  \multicolumn{1}{r|}{31.3 / 33.8} &
  21.4 / 32.1 \\ \hline
\texttt{ASR-multi} / \texttt{MTASS} + ${L}_{dis}$ &
  \multicolumn{1}{r|}{9.3 / 27.3} &
  \multicolumn{1}{r|}{13.0 / 28.3} &
  \multicolumn{1}{r|}{24.5 / 30.8} &
  \multicolumn{1}{r|}{29.1 / 33.9} &
  \multicolumn{1}{r|}{31.5 / 31.4} &
  21.5 / 30.3 \\ \hline
\texttt{ASR-multi} / \texttt{MTASS} + ${L}_{dis}$ + ${L}_{cst}$ &
  \multicolumn{1}{r|}{9.0 / 25.2} &
  \multicolumn{1}{r|}{12.7 / 26.5} &
  \multicolumn{1}{r|}{23.8 / 29.1} &
  \multicolumn{1}{r|}{28.4 / 31.9} &
  \multicolumn{1}{r|}{30.6 / 29.2} &
  20.9 / 28.4 \\ \hline
\end{tabular}

\caption{The CERs of the JRSV on the test sets with different overlap ratios.}
  \label{tab:jrsv_exp}
  \centering
  \vspace{-10pt}
\begin{tabular}{|l|rrrrrr|}
\hline
 & \multicolumn{6}{c|}{CER \% (Speech / Singing Voices)} \\ \hline
\multicolumn{1}{|c|}{Overlap Ratios} &
  \multicolumn{1}{c|}{0.0} &
  \multicolumn{1}{c|}{0.1} &
  \multicolumn{1}{c|}{0.3} &
  \multicolumn{1}{c|}{0.5} &
  \multicolumn{1}{c|}{1.0} &
  \multicolumn{1}{c|}{Avg.} \\ \hline\hline
\texttt{JRSV-t} (trainable MTASS) &
  \multicolumn{1}{r|}{10.1 / 22.0} &
  \multicolumn{1}{r|}{12.1 / 22.6} &
  \multicolumn{1}{r|}{17.9 / 25.2} &
  \multicolumn{1}{r|}{20.2 / 26.7} &
  \multicolumn{1}{r|}{20.4 / 28.1} &
  16.1 / 24.9 \\ \hline
\texttt{JRSV-f} (frozen MTASS) &
  \multicolumn{1}{r|}{8.2 / 13.1} &
  \multicolumn{1}{r|}{9.7 / 13.8} &
  \multicolumn{1}{r|}{14.1 / 14.7} &
  \multicolumn{1}{r|}{15.9 / 15.5} &
  \multicolumn{1}{r|}{16.6 / 16.5} &
  12.9 / 14.7 \\ \hline
\texttt{JRSV-f-d} (frozen MTASS + $L_{\text{distil}}$) &
  \multicolumn{1}{r|}{\textbf{7.6 / 10.8}} &
  \multicolumn{1}{r|}{\textbf{9.3 / 11.6}} &
  \multicolumn{1}{r|}{\textbf{13.6 / 12.0}} &
  \multicolumn{1}{r|}{\textbf{15.3 / 12.8}} &
  \multicolumn{1}{r|}{\textbf{15.9 / 13.1}} &
  \textbf{12.3 / 12.1} \\ \hline
\end{tabular}
\vspace{-10pt}
\end{table*}

\subsection{Preliminary experiment: directly recognizing the mixture}
We evaluate the performance of the ASR model without MTASS to show the impacts of the complex acoustic conditions on ASR. The results are shown in \autoref{tab:pri_exp}. \texttt{ASR-speech}, \texttt{ASR-sing}, and \texttt{ASR-multi} denote the ASR model trained on the pure speech data, pure singing data, and both speech and singing data, respectively. We can see that the models trained on the matched data can perform well on the corresponding unmixed data. \texttt{ASR-multi} performs best on the unmixed speech and singing voices. However, all three models fail to recognize the mixed audio. In particular, the CERs of the singing voice are larger than 100\%. The reason is that many speech contents are recognized and inserted, which causes insertion errors in the results of singing voices. This experiment demonstrates that the ASR model fails to recognize mixed audio in complex acoustic conditions directly.

\subsection{Recognizing with the cascade system}
We evaluate the performance of the cascade system. Namely, separate the mixed audio into the speech track and the singing voice track, and then recognize with \texttt{ASR-multi}, achieving the best performance for the unmixed data. 

First, we evaluate the performance of the MTASS models in \autoref{tab:sep_exp}. We can see that \texttt{MTASS} achieves a significant SDR improvement. The discriminative loss ${L}_{dis}$ and the consistent loss ${L}_{cst}$ can further improve the SDR. Then we evaluate the recognition performance in \autoref{tab:cascade_exp}. ${L}_{dis}$ brings an improvement for sing voices. When using both ${L}_{dis}$ and ${L}_{cst}$, the system achieves the best performance.

\subsection{Recognizing with JRSV}
\label{subsec:exp_jrsv}
We evaluate the effectiveness of the proposed JRSV. First, comparing \autoref{tab:jrsv_exp} with \autoref{tab:cascade_exp}, we can see that \texttt{JSRV-f} with the frozen MTASS achieves significantly better performances than \texttt{ASR-multi} / \texttt{MTASS} cascade models. With online distillation, the performance is further improved. On average, the \texttt{JRSV-f-d} achieves a 41\%/57\% relative CER reduction compared with the best cascade system.

Counter-intuitively, \texttt{JRSV-t} (trainable MTASS) does not achieve a good performance. We have searched many weights for $L_{\text{MTASS}}$ and $L_{\text{ASR}}$ but does not achieve a positive result. We analyze a possible reason that the MTASS module and the ASR module play different roles: the MTASS module processes the low-level features, and the ASR module processes the high-level semantic representations. The two objects are not compatible. The ASR loss influences the low-level feature extraction, which affects the performance.

\section{Conclusions and future work}
We propose JRSV to jointly recognize speech and singing voices. The MTASS module separates the mixed audio into distinct speech and singing voice tracks while removing background music. The CTC/attention hybrid recognition module recognizes both tracks. Online distillation is proposed to further improve recognition accuracy. To evaluate the proposed methods, a benchmark dataset is constructed and will be released. Experimental results demonstrate that this methods can significantly improve recognition accuracy on each track of the mixed audio. In the future, we will investigate why the joint optimization of $L_{\text{MTASS}}$ and $L_{\text{ASR}}$ does not bring better performance in more detail.

\bibliographystyle{IEEEtran}
\bibliography{mybib}

\end{document}